\documentclass[12pt,a4paper]{article}
\usepackage[utf8]{inputenc}
\usepackage{amsmath}
\usepackage{amsfonts}
\usepackage{amssymb}
\usepackage{amsbsy}
\usepackage{graphicx}
\usepackage[sharp]{easylist}
\newtheorem{theorem}{Theorem}

\newtheorem{corollary}{Corollary}

\begin{document}
\title{PDF Steganography based on Chinese Remainder Theorem}
\author{Ren\'{e} Ndoundam, St\'{e}phane Gael R. Ekodeck \\
{\small University  of Yaounde I, LIRIMA, Team  GRIMCAPE, P.o.Box  812 Yaounde, Cameroon}  \\
{\small IRD, UMI 209, UMMISCO, IRD France Nord, F-93143, Bondy, France; }  \\
{\small Sorbonne Unversit\'es, Univ. Paris 06, UMI 209, UMMISCO, F-75005, Paris, France} \\
{\small E.mail : ndoundam@gmail.com , ekodeckstephane@gmail.com} 
 }
\date{}
\maketitle {}
\begin{abstract} 
We propose different approaches of PDF files based steganography, essentially based on the Chinese Remainder Theorem. Here, after a cover PDF document has been released from unnecessary characters of ASCII code $A0$, a secret message is hidden in it using one of the proposed approaches, making it invisible to common PDF readers, and the file is then transmitted through a non-secure communication channel. Where each of our methods, tries to ensure the condition that the number of inserted $A0$ is less than the number of characters of the secret message $s$.
\end{abstract}
{\bf Keywords:}  Steganography, PDF files and readers, Chinese Remainder Theorem.

\section{Introduction}

Steganography consists in hiding a secret message in public document acting as a covert, in a way that sent through a non-secure communication channel, only the sender and the receiver are able to understand it, and anyone else cannot distinguish the existence of an hidden message. It is one of the Information hiding techniques as showed on figure 1, 
\begin{figure}[h]
\centering
\includegraphics[scale=0.45]{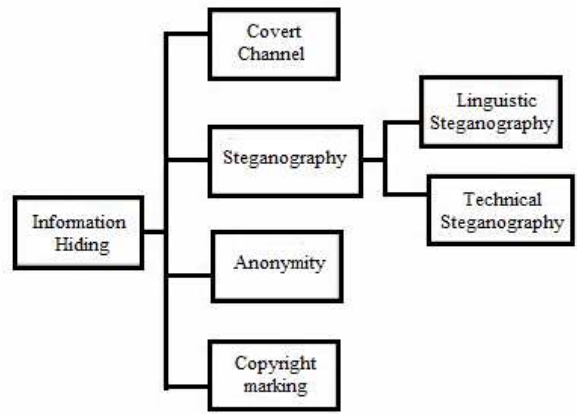} 
\caption{Classification of information hiding techniques}
\end{figure}
where Linguistic Steganography is defined by Chapman \textit{et al} \cite{PD 08} as,\textit{ “the art of using written natural language to conceal secret messages”},  and Technical Steganography is defined as a structure rather than a text, that can be represented by any physical means such as invisible inks, microdots \cite{PD 08}.
Most of the work in steganography has been done on images, video clips, music, sounds and texts. But, text steganography is the most complex, due to the lack of redundant information in text files, whereas lot of redundancy is present in image or sound files, leading to a high exploitation of those files in steganography \cite{GKDS 12}. \\
There are several approaches encountered in the literature regarding the text steganography such as, line shift, word shift, syntactic methods, etc. Subsequently we focused on the steganography based on PDF files.

\section{PDF files based Steganography}

PDF, created by Adobe Systems \cite{ASI 06} for document exchange, is a fixed-layout format for representing documents in a manner independent of the application software, hardware, and operation system. PDF files are frequently used nowadays and this fact makes it possible to use them as cover documents in information hiding. Studies using these files as cover media, are very few. \\
Our proposal is based on the work of \textit{I-Shi et al.} \cite{IW 10}, in which secret data are embedded at between-word or between-character locations in a PDF file, by using the non-breaking space with American Standard Code for Information Interchange (ASCII) code \textit{A0}. \textit{I-Shi et al.} \cite{IW 10} found in their study that, the non-breaking space (\textit{A0}) is a character when embedded in a string of text characters, becomes invisible in the windows of several versions of common PDF readers, and use that phenomenon for data hiding. They showed two types of invisibility, based on the ASCII code \textit{A0}. \\

The first one is created by specifying the width of \textit{A0} appearing in the PDF reader's window to be the same as that of the original white-space represented by the ASCII code \textit{20}. The width of an ASCII code, is the width of the character represented by the code as displayed in a PDF reader's window. Subsequently, \textit{A0} and \textit{20} become white-spaces. Their approach based on this first  type of invisibility called \textit{alternative space coding}, uses \textit{A0} and \textit{20} in a PDF text alternatively as a between-word space to encode a message bit b according to the following binary coding technique:
\begin{tabbing}
\hspace*{.25in}\=\hspace{2ex}\=\hspace{2ex}\=\hspace{2ex}\kill
\> \> \> if b = 1; then replace \textit{20} between two words by \textit{A0};\\
\> \> \> if b = 0; make no change.
\end{tabbing}

This approach has the advantage of incurring no increase of the PDF file size because it just replaces the space exhibited by the code \textit{20} by another exhibited by the code \textit{A0}. However, if the between-word locations in a PDF page are few, then only a small number of bits may be embedded.\\

The second one is created by setting the width of the ASCII code \textit{A0} to be \textit{zero} in a PDF page. They found in their study an \textit{A0} doesn't appear in a PDF reader's window just like if it was nonexistent. Their approach called \textit{null space coding}, given a \textit{message character C}, embeds it at a location \textit{L} as follows:
\begin{center}
 if the index of \textit{C} as specified in \textit{Table 1} is \textit{m},\\
   then embed \textit{m} consecutive \textit{A0's} at location \textit{L}.\\
\end{center}

In this approach they presented,  \textit{Table 1} \cite{IW 10} contains ASCII codes selected for message representations in their study, each one indexed with an integer value.\\

The advantage of this approach is that the number of between-character locations are higher than the between-word locations. This makes the efficiency of the \textit{null space coding} much higher. But an obvious disadvantage is that the resulting PDF file size will be higher than the original one (the one without \textit{A0's} embedded in it).\\

Our work is based on this last type of invisibility described by \textit{I-Shi et al.}, such that our problematic is to reduce the weight difference between the cover PDF file and the stego PDF file resulting from the embedding process, while increasing the embedding capacity of the cover PDF file. In order to reduce considerably the risks of detecting a cover communication based on the file size.

\section{Our Contribution}

Given a secret message $s$ to be conceal in a cover text message, the null space coding developed by \textit{I-Shi et al.}, proceeds as follows:
\begin{itemize}
\item Firstly, $s$ is compressed using the Huffman coding, where at the end a file, containing a table where each line has a letter of $s$ followed by a value, is generated;
\item Secondly, for each character of $s$ a number of $A0$'s is inserted in the cover text equivalent to the value generated by the Huffman coding for that character, thus producing a stegotext.
\item Thirdly, the file and the stegotext are transmitted through a non-secure communication channel. We note that two files (the file containing Huffman codes for the characters of the secret message and the PDF file resulting from the embedding method) are transmitted.
\end{itemize}

Their method cannot guarantee that the number of embedded $A0$'s is less than the number of characters of $s$ or at least if $s$ grows higher, the number of inserted $A0$'s won't explode.\\

Our aim is to propose different approaches, based on the Chinese Remainder Theorem, which their goal is to attain the above conditions and transmit one and only one file (more precisely only the stegotext), through a non-secure communication channel.

\section{Chinese Remainder Theorem}
\begin{theorem}
Let $\{n_i\}_{i=1}^k$ be a pairwise relatively prime family of positive integers, and let $a_1 , ... , a_k$ be arbitrary integers. Then there exists a solution $x \in \mathbb{Z}$ to the system of congruence
\begin{center}
$	\begin{cases}
		x \equiv a_1 \mod n_1 \\
		x \equiv a_2 \mod n_2\\
		\cdots \\
		
		x \equiv a_k \mod n_k\\
\end{cases}$
\end{center}
Moreover, any $a' \in \mathbb{Z}$ is a solution to this system of congruence if and only if $a \equiv a' (mod\ N)$, where $N = \prod_{i = 1}^k n_i$
\end{theorem}
$\blacksquare$\\

Given $a_i$ and $n_i$, $\ (1 \leq i \leq k$), we present the classic method of construction of $x$ from $a_i$ and $n_i$ as follows:\\

We first construct integers $e_i$,  $(1 \leq i \leq k)$, such that for $i,\ j = 1,\cdots ,k$, we have:
\begin{gather} \label{eeq:n1}
e_j \equiv
\begin{cases}
		1 \mod n_i, \ if \ j = i \\
		0 \mod n_i, \ if \ j \neq i \\
\end{cases}
\end{gather}
Then setting 
\begin{center}
$x = \sum_{i = 1}^k a_i e_i$
\end{center}
Allows to see that for $j = 1, \cdots, k$ we have
\begin{center}
$x \equiv \sum_{i = 1}^k a_i e_i \ \equiv a_j \mod n_j$
\end{center}
As all the terms in this sum are zero modulo $n_j$, except for the term $i = j$, which is congruent to $a_j \mod n_j$. To construct $e_i$,  $(1 \leq i \leq k)$, satisfying (1), let us define $b_i = N/n_i$, which is the product of all the moduli $n_j$ with $j \neq i$. Then, $c_i$ and $e_i$ are defined as follows: $c_i = (b_i)^{-1} \mod n_i$ and $e_i = b_i c_i$.\\

Garner's algorithm is an efficient method for determining $x$, $0 \leq a < N$, given $a(x)$ = $(a_1, a_2,..., a_k)$, the residues of $x$ modulo the pairwise co-prime moduli $n_1, n_2, ..., n_k$ \cite{MOV 96}. \\

\textbf{Garner's algorithm for CRT} \cite{MOV 96}\\
\textbf{Input:} a positive integer $M = \prod_{i = 1}^t m_i > 1$, with $gcd(m_i, m_j) = 1$ for all $i \neq j$, and a modular representation $v(x) = (v_1, v_2, \cdots, v_t)$ of $x$ for the $m_i$.\\
\textbf{Output:} the integer $x$ in radix $b$ representation.
\textit{
\begin{tabbing}
1.  For \= $i$ from 2 to $t$ do the following: \+ \\
1.1. 	$C_i \leftarrow 1.$ \\
1.2. 	For $j$ \= from 1 to $(i-1)$ do the following:\+ \\
			$u \leftarrow m_j^{-1} mod\ m_i$\\
			$C_i \leftarrow u \times C_i\ mod\ m_i$\- \- \\
2. 	$u \leftarrow v_1, x\leftarrow u.$	\\
3. For \= $i$ from 2 to $t$ do the following:\+ \\
 $u \leftarrow (v_i - x) \times C_i\ mod\ m_i, x \leftarrow x + u \times \prod_{j=1}^{i-1} m_j$\- \\
4. Return(x).
\end{tabbing}
}
\textbf{Time Complexity}: $O(n^2)$\\
This theorem is highly useful in a many contexts as, randomized primality test, modular arithmetic, secret sharing, etc.

\section{Preprocessing on the cover file}

The PDF file $f \in F$, that would be used as cover, needs to be cleansed of all $A0$'s contained in it. Meaning, going from the beginning of the file to its end, if we cross a $A0$ with size different from $0$, we replace it by a space character (ASCII code 20), and if we cross a $A0$ of size $0$, we remove it, as presented by the following function.\\
\textbf{Input:}  f: cover PDF file \\
\textbf{Output:} f: cover PDF file with no sequence of more than one \textit{A0}
\textit{
\begin{tabbing}
1. Open the file $f$;\\
2.		Browse the PDF file $f$ character by character  and \\
\ \ \ 	fo\= r each encountered $A0$ do: \+ \\
2.1 		If ($sizeof(A0) > 0$) then replace $A0$ by a space character;\\
2.2			else remove $A0$ from $f$;\-  \\				
3.	Save and close the file $f$;\\	
4.	Return $f$;
\end{tabbing}
}
Where, $sizeof(A0)$ is a function that retrieves the width of the non-breaking space character, if exists, set in a cover PDF file $f$.\\
\textbf{Time Complexity}: $O(|f|)$\\

The reason why we apply this procedure on a cover PDF file, is to ensure that the file has not been modified by a steganographic technique based on ASCII code $A0$; and also, as $A0$ by default has the width of the space character, it can be replaced by it, all this to avoid ambiguity between $A0$ inserted by our techniques and those found initially in the cover file.

\section{Presentation of the different approaches}
For the sender and the receiver to be able to communicate through a non-secure channel, they have to agree on a secret key that would be use to encrypt a secret message, that would be send one to another. Regarding our approaches, the key $k \in \mathbb{N}$, represents the number of bits (block length) in which a secret message $s \in \{0,1\}^*$ would be split into before its encoding. And it's previously selected by the sender and the receiver and shared through a secure channel. Subsequently $|s|$ denotes the length of the string $s$.

\subsection{First Approach}
\subsubsection{Hiding method}
We denote $s$ the secret message, an integer $k$ a secret key and $f$ a cover PDF file. Without loss of generality, we assume that the length of $s$ is a multiple of $k$. The first approach proceeds as follows: \\
\textbf{Input:}  s: secret message; k: secret key; f: cover PDF file. \\
\textbf{Output:} f: cover PDF file with embedded \textit{A0}'s\\

\textbf{Step 1}: two co-primes $p_1, p_2$, are computed from $k$ such that,
\begin{center}
$p_1 = 2^{\lceil{\frac{k}{2}}\rceil}$; $p_2 = p_1 + 1$.
\end{center} 

\textbf{Step 2}: $s$ is split in $n$ blocks of length $k$ stored the matrix $sp$ such that:
\begin{center}
 $sp[i, j] = s[(i - 1)k + j],\ 1 \leq i \leq n, 1 \leq j \leq k$.
\end{center}

\textbf{Step 3}: each line of $sp$ corresponding to a binary sequence, is transformed in its decimal value $dec[i]$ such that,
\begin{center}
$dec[i] = \sum_{j = 1}^k sp[i, k-j+1] \times 2^{(j-1)},\ 1 \leq i \leq n$.
\end{center}

\textbf{Step 4}:  for each decimal value $dec[i]$ $(1 \leq i \leq n)$, two remainders $r[1,i]$ and $r[2,i]$, are computed such that 
\begin{center}
$r[1,i] = dec[i]\ mod\ p_1 $ and $ r[2,i] = dec[i]\ mod\ p_2,\ 1 \leq i \leq n$
\end{center} 

\textbf{Step 5}: each $r[j,i], (1 \leq j \leq 2 \ and\ 1 \leq i \leq n)$, obtained from the previous step is transformed in its binary value stored in a matrix $binr$ bit by bit, such that:
\begin{center}
$binr[((i-1)\times 2 + j),1] \cdots  binr[((i-1)\times 2 + j),\lceil{\frac{k}{2}}\rceil] = binDecomp(r[j,i], \lceil{\frac{k}{2}}\rceil)$,\\
$1 \leq j \leq 2 \ and\ 1 \leq i \leq n$
\end{center}
Where, $binDecomp(r[j,i], \lceil{\frac{k}{2}}\rceil)$ is a function that returns the binary decomposition of a remainder $r[j,i]$ on $\lceil{\frac{k}{2}}\rceil$ bits of length.\\

\textbf{Step 6}: Add a column at $binr$, the number of columns would then move from $\lceil{\frac{k}{2}}\rceil$ to $(1 + \lceil{\frac{k}{2}}\rceil)$; and for each line add a control bit at the end as shown by the following:
\textit{
\begin{tabbing}
1.	for (i := 1 to (2$\times$n - 1)) do 	binr[i, ($1 + \lceil{\frac{k}{2}}\rceil$)] := 0;\\
2. binr[2n, ($1 + \lceil{\frac{k}{2}}\rceil$)] := 1;
\end{tabbing}
}

\textbf{Step 7}: each line of $binr$ is embedded in a cover PDF file $f$, as described by the following:
\textit{
\begin{tabbing}
1.	Get the first between-character location $lc$;\\
2.	fo\= r (i := 1 to 2$\times$n) do\+ \\
2.1.	fo\= r (j := 1 to (1 + $\lceil{\frac{k}{2}}\rceil$)) do\+ \\
			be\= gin\+ \\
2.1.1.		if $(binr[i,j] = 1)$ then Insert $A0$ at $lc$ in the file $f$;\\
2.1.2.		Get the next between-character location $lc$\- \\
			end;\\
\end{tabbing}
}

The control bit is there to help, during the recovery procedure, to know when to stop looking for embedded blocks in the cover file.\\
\textbf{Time Complexity}: $O(n*k)$

\subsubsection{Recovery method}
To recover secret message from a stego PDF file encoded with the above procedure, the binary sequences encoded with $A0$'s in the file must be recover at first, then remainders that produced those sequences, and with the $k$, computer the values related to those remainders, as described by the following procedure:\\
\textbf{Input:} f: stego-PDF file, k: secret key \\
\textbf{Output:} s: secret message\\

\textbf{Step 1}:two co-primes $p_1, p_2$, are computed from $k$ such that,
\begin{center}
$p_1 = 2^{\lceil{\frac{k}{2}}\rceil}$; $p_2 = p_1 + 1$.
\end{center} 

\textbf{Step 2}: retrieve the different lines of $binr$ as follows:
\textit{
\begin{tabbing}
1.	i := 1; \\
2.	exist := true; \\
3.	n := 0;\\
4.	Get the first couple of characters (a, b) from $f$;\\
5.	wh\= ile (exist and !feof(f)) do\+ \\
		be\= gin\+ \\
5.1.		j := 1;\\
5.2.		wh\= ile (j $\leq$ (1 + $\lceil{\frac{k}{2}}\rceil$)) do\+ \\
				be\= gin\+ \\
				if (a != A0 and b != A0) then binr[i,j] := 0;\\
				el\= se\+ \\ 
					if (a!= A0 and b = A0) then binr[i,j] := 1; \\
					el\= se\+ \\ 
						if (a = A0 and b = A0) then exist = false; \\
						else j := j - 1; \\
						endif;\- \\
					endif;\- \\
				endif\\
				j := j + 1;\\
				c := the next character in f;\\
				a := b;\\
				b := c; \- \\
				end;\- \\
5.3.		i := i +1;\- \\
		end;\- \\
6.	n := i - 1;
\end{tabbing}
}

\textbf{Step 3}: remove from $binr$ the $(1 + \lceil{\frac{k}{2}}\rceil)^{th}$ column, corresponding to the control bit's column.\\

\textbf{Step 4}: compute each $r[j, i], (1 \leq j \leq 2 \ and\ 1 \leq i \leq n)$ from each line of $binr$ such that:
\begin{center}
$r[j, i] = \sum_{l = 1}^{\lceil{\frac{k}{2}}\rceil} binr[i, \lceil{\frac{k}{2}}\rceil-l+1] \times 2^{(l-1)},\ 1 \leq i \leq n$.
\end{center}

\textbf{Step 5}: compute each $dec[i], 1 \leq i \leq n)$ using Garner's algorithm such that:
\begin{center}
$dec[i] = GarnerAlgorithm(\{p_1,p_2\}, \{r[1, i], r[2, i]\})$\\
$1 \leq i \leq n$.
\end{center}

\textbf{Step 6}: transform each $dec[i]$ in its binary sequence $sp[i,j],\ (1 \leq j \leq k)$ bits such that:
\begin{center}
$(dec[i])_2  =  \underbrace{sp[i, 1] sp[i, 2]\cdots sp[i, k]}_{k\ bits}$
\end{center}

\textbf{Step 7}: merge all the binary string into one, the secret $s$, such that:
\begin{center}
$ s[(i - 1)k + j] = sp[i, j],\ 1 \leq i \leq n, 1 \leq j \leq k$.
\end{center}

Where $GarnerAlgorithm$ take as input a list of co-primes $p_1,p_2$, a list of remainders $r[1, i], r[2, i]$, and outputs a unique value $dec[i]$.\\
\textbf{Time Complexity}: $O(n*k)$

\subsubsection{Evaluation}
	In this approach, for each block of length $k$, 2 remainders $r_1$, $r_2$ are computed respectively from $p_1$ and $p_2$. As $p_1 < p_2$, we can easily deduce that, the number max of inserted $A0$'s from a remainder is:
\begin{center}
$log_2(Max(r_1, r_2)) = log_2(p_1) = {\lceil \frac{k}{2} \rceil}$.
\end{center}
	Thus, the number max of $A0$'s that can be inserted for a block of $s$ is $k$.\\
	
	So, to embed a full secret message $s$ divided into $n$ blocks of length $k$, the maximum number of $A0$'s that would be needed is:
\begin{center}
 $n*k + 1 = |s| + 1 > |s|$.
\end{center} 
  We add $1$ here because, for the last computed remainder, a A0 character would be inserted at the end of the hiding procedure, serving as ending point for the recovery method. From, all these comes out the following theorem.
\begin{theorem}
.\\
Given a secret message $s$, a secret key $k$ such that number of blocks of length $k$, is given by $n = \frac{|s|}{k}$, and two primes $p_1, p_2$ such that $p_1 = 2^{\lceil{\frac{k}{2}}\rceil}$, $p_2 = p_1 + 1$, the number $N$ of $A0$'s insertions at between-character locations to perform in a PDF file, is:
\begin{center}
$N \leq |s|+1$
\end{center}

\end{theorem}

$\blacksquare$\\

Where $N$ depends on the number of bits having value 1, contained in the secret message $s$'s computed remainders.

\subsection{Second Approach}
\subsubsection{Hiding method}
In this particular approach, what would be considered as key is not $k$ the block length, but $m$, a value that allows to compute primes between $2 \times m$ and $3 \times m$, such that the base 2 logarithm of the product of all those primes gives us the block length $k$, in which a secret message $s$ would be divided in. Those primes allows us to compute remainders, which their values would be used to compute position where one $A0$ would inserted. The whole procedure is defined as follows:\\
\textbf{Input:} s: secret message, m: secret key, f: cover PDF file \\
\textbf{Output:} f: stego-PDF file\\

\textbf{Step 1}: compute primes $p_1,p_2,\cdots p_t$ such that:
\begin{center}
$2 \times m \leq p_1 < p_2 < \cdots < p_t \leq 3 \times m$
\end{center}
where, $t$ is the number of primes computed between $2 \times m$ and $3 \times m$.

\textbf{Step 2}: compute the block length $k$ such that:
\begin{center}
$k = \lfloor{log_2(prod)}\rfloor$
\end{center}
where, $prod = \prod_{i = 1}^t p_i$.

\textbf{Step 3}: $s$ is split in $n$ blocks of length $k$ stored the matrix $sp$ such that:
\begin{center}
 $sp[i, j] = s[(i - 1)k + j],\ 1 \leq i \leq n, 1 \leq j \leq k$.
\end{center}

\textbf{Step 4}: each line of $sp$ corresponding to a binary sequence, is transformed in its decimal value $dec[i]$ such that,
\begin{center}
$dec[i] = \sum_{j = 1}^k sp[i, k-j+1] \times 2^{(j-1)},\ 1 \leq i \leq n$.
\end{center}

\textbf{Step 5}: for each decimal value $dec[i]$ $(1 \leq i \leq n)$, remainders $r[1,i], r[2,i],\cdots, r[t,i]$, are computed such that 
\begin{center}
$r[j,i] = dec[i]\mod p_j,\ 1 \leq i \leq n,\ 1 \leq j \leq t$
\end{center} 

\textbf{Step 6}: for each remainder $r[j,i],\ 1 \leq i \leq n,\ 1 \leq j \leq t$, we compute positions $pos[1,1],\cdots, pos[t,n]$, as described by the following procedure:
\textit{
\begin{tabbing}
1.	l := 0; i := 1; n := dec.length; h := $t \times p_t$;\\
2.	wh\= ile (i $\leq$ n) do\+ \\	
		be\= gin\+ \\
2.1.		fo\= r (j := 1 to t) do\+ \\
				pos[$(t \times (i - 1)) +j$] := l + (j - 1) + $(t \times r[j,i])$;\- \\
2.2.		l := l + h; \\
2.3.		i := i + 1;\- \\
		end;
\end{tabbing}
}

\textbf{Step 7}: sort the vector $pos$ in the ascending order;\\

\textbf{Step 8}: for each $pos[i], 1 \leq i \leq (n\times t) - 1$, insert one $A0$ at the $pos[i]^{th}$ between-character location of $f$. And, at the $pos[n\times t]^{th}$ between-character location of $f$, insert two $A0$'s, to mark then end of the process.\\
\textbf{Time Complexity}: $O(n*k)$

\subsubsection{Recovery method}
To recover secret message from a stego PDF file encoded with the above procedure, the positions of all the $A0$'s in the file must be recover at first, then remainders that produced those positions, and with the $k$, computer the values related to those remainders, as described by the following procedure:\\
\textbf{Input:} f: stego-PDF file, m: secret key \\
\textbf{Output:} s: secret message\\

\textbf{Step 1}: compute primes $p_1,p_2,\cdots p_t$ such that:
\begin{center}
$2\times m \leq p_1 < p_2 < \cdots < p_t \leq 3\times m$
\end{center}

\textbf{Step 2}: compute the block length $k$ such that:
\begin{center}
$k = \lfloor{log_2(prod)}\rfloor$
\end{center}
where, $prod = \prod_{i = 1}^t p_i$.

\textbf{Step 3}: compute the block length in the file $f$ such that:
\begin{center}
$h = t \times p_t$
\end{center}

\textbf{Step 4}: retrieve the positions where $A0$'s have been inserted as described below:
\textit{
\begin{tabbing}
1.	i := 1; count := 1; n := 0; exist := true;\\
2.	get the first couple (a, b) of characters from $f$;\\
3.	wh\= ile (exist and !feof(f)) do\+ \\	
		be\= gin\+ \\
			if\=  (a != $A0$ and b = $A0$) then \+ \\
					be\= gin\+ \\
						pos[$i$] := count;\\
						i := i + 1;\- \\
					end; \- \\
			el\= se\+ \\ 
				if (a != $A0$ and b != $A0$) then do nothing; \\
				el\= se\+ \\ 
					if (a = $A0$ and b != $A0$) then count := count - 1;\\
					el\= se\+ \\
						if (a = $A0$ and b = $A0$) then exist := false;\\
						endif;\- \\
					endif;\- \\
				endif;\- \\
			endif;\\
			count := count + 1;\\
			c := the next character in f;\\
			a := b; \\
			b := c;\- \\
		end;\- \\
4.	n := (i - 1) / t; 
\end{tabbing}
}

\textbf{Step 5}: Compute the remainders from the table $pos$ as follows:
\textit{
\begin{tabbing}
1.	l = 0; n = pos.length / t;\\
2.	fo\= r (i := 1 to n) do\+ \\	
		be\= gin\+ \\
2.1.		fo\= r (j := 1 to t) do\+ \\
				f := $[pos[(t \times (i - 1)) + j] - (l + j - 1)] \mod t$;\\
				r[f, i] := $[pos[(t \times (i - 1)) + f] - (l + f - 1)] / t$;\- \\
2.2.		l := l + h; \- \\
		end;
\end{tabbing}
}

\textbf{Step 6}: Compute each decimal value $dec[i]$ of a block of $s$ such that:
\begin{center}
$dec[i] = GarnerAlgorithm(\{p_1,p_2,\cdots p_t\}, \{r[1, i], r[2, i],\cdots, r[t, i]\})$\\
$1 \leq i \leq n$.
\end{center}

\textbf{Step 7}: transform each $dec[i]$ in its binary sequence $sp[i,j],\ (1 \leq j \leq k)$ bits such that:
\begin{center}
$(dec[i])_2  =  \underbrace{sp[i, 1] sp[i, 2]\cdots sp[i, k]}_{k\ bits}$
\end{center}

\textbf{Step 8}: merge all the binary string into one, the secret $s$, such that:
\begin{center}
$ s[(i - 1)k + j] = sp[i, j],\ 1 \leq i \leq n, 1 \leq j \leq k$.
\end{center}

Where $GarnerAlgorithm$ take as input a list of co-primes $p_1,p_2,\cdots p_t$, a list of remainders $r[1, i], r[2, i],\cdots, r[t, i]$, and outputs a unique value $dec[i]$.\\
\textbf{Time Complexity}: $O(n*k)$

\subsubsection{Evaluation}
Let:
\begin{itemize}
\item $\vartheta(x)= \sum_{p \leq x,\ p\ prime} ln(p)$,\\
\item $\pi (x)$ the number of prime numbers less or equal to $x$,\\
\item $p_1,p_2,\cdots p_t$ are the prime numbers taken between $2m$ and $3m$.\\
\end{itemize}

In this approach, for each block of length $k$, $t\ A0$'s are inserted in the cover file. So to embed a full secret message $s$ divided into $n$ blocks of length $k$, $tn$ $A0$'s would be needed. This is the result we obtained, resume by the following theorem. Regardless the number of blocks we need to embed, an additional $A0$, would be added to allow the recovery method to stop when all the hidden bits have been recovered.

\begin{theorem}
.\\
Given a secret message $s$, a secret key $m$, a set of primes $p_1,p_2,\cdots p_t$ taken between $2m$ and $3m$, $k$ the block length such that $k = \lfloor{log_2 \prod_{i = 1}^t p_i}\rfloor$, and $n$ the number of blocks of length $k$, such that $n = \lceil \frac{|s|}{k} \rceil$. The number $N$ of $A0$'s insertions at between-character locations, to perform in a PDF file is given by:
\begin{center}
$N =
\begin{cases}
		t+1, \ if \ |s| \leq k \\
		(t*n)+1, \ if \ |s| > k \\
\end{cases}$
\end{center}

\end{theorem}
$\blacksquare$\\

On one hand, as $t$ is the number of primes taken between $2m$ and $3m$,
\begin{center}
$t = \pi(3m) - \pi(2m)$ 
\end{center}
And from the work of Hadamard and de la Vallée Poussin \cite{LPV 00}, which  resulted in the following theorem:\\
\textbf{The Prime Number Theorem \cite{LPV 00}:}\\

\textit{
Let $\pi(n)$ denote the the number of primes among 1, 2, $\cdots$, n. Then,
	\begin{center}
			$\pi(n) \sim \frac{n}{ln(n)}$
	\end{center}
}
$\blacksquare$\\

 We can deduce that:
	\begin{equation} \label{E:mm8} 
	t \sim \frac{3m}{ln(3m)} - \frac{2m}{ln(2m)}
	\end{equation}

On the other hand, from estimations of Rosser and Schoenfeld \cite{ROS 62}, we have:\\
\begin{center}
$ 
\begin{cases}
\vartheta(x) < x(1 + \frac{1}{2ln(x)}),\ for\ 1 < x \leq 41 \\
\vartheta(x) > x(1 - \frac{1}{ln(x)}),\ for\ 41 < x\\
\end{cases}
$
\end{center}

We can deduce that:
\begin{center}
$x - \frac{3x}{ln(3x)} - \frac{2x}{2ln(2x)} < \vartheta(3x) - \vartheta(2x) < x + \frac{3x}{2ln(3x)} + \frac{2x}{ln(2x)}$.
\end{center}

It is easy to show that $\forall x \in \mathbb{R},\ x \geq e^5$ we have:
\begin{center}
$x - \frac{3x}{ln(3e^5)} - \frac{2x}{2ln(2e^5)} \leq \vartheta(3x) - \vartheta(2x) \leq x + \frac{3x}{2ln(3e^5)} + \frac{2x}{ln(2e^5)}$
\end{center}

From these estimations, we deduce that, for $x \geq e^5$:
\begin{center}
$\frac{2}{10}x \leq \vartheta(3x) - \vartheta(2x) \leq \frac{17}{10}x$.
\end{center}

Thus, putting $m = x$:
\begin{equation}  \label{E:mm9}
\frac{2}{10}m \leq k \leq \frac{17}{10}m.
\end{equation}

From (\ref{E:mm8}) and (\ref{E:mm9}), we can deduce the following corollary.
\begin{corollary}
.\\$\forall m \geq e^5$, the number $N$ of $A0$'s insertions at between-character locations, to perform in a PDF file is given by:
\begin{center}
$N \sim
\begin{cases}
		\frac{3m}{ln(3m)} - \frac{2m}{ln(2m)} + 1, \ if \ |s| \leq \frac{17}{10}m \\
		(\frac{3m}{ln(3m)} - \frac{2m}{ln(2m)}) * n + 1, \ if \ |s| > \frac{17}{10}m \\
\end{cases}$
\end{center}
\end{corollary}

\subsection{Third Approach}
\subsubsection{Hiding method}
\textbf{Input:}  s: secret message; k: secret key; f: cover PDF file. \\
\textbf{Output:} f: cover PDF file with embedded \textit{A0}'s\\

\textbf{Step 1}: two co-primes $p_1, p_2$, are computed from $k$ such that,
\begin{center}
$p_1 = 2^{\lceil{\frac{k}{2}}\rceil}$; $p_2 = p_1 + 1$.
\end{center} 

\textbf{Step 2}: $s$ is split in $n$ blocks of length $k$ stored the matrix $sp$ such that:
\begin{center}
 $sp[i, j] = s[(i - 1)k + j],\ 1 \leq i \leq n, 1 \leq j \leq k$.
\end{center}

\textbf{Step 3}: each line of $sp$ corresponding to a binary sequence, is transformed in its decimal value $dec[i]$ such that,
\begin{center}
$dec[i] = \sum_{j = 1}^k sp[i, k-j+1] \times 2^{(j-1)},\ 1 \leq i \leq n$.
\end{center}

\textbf{Step 4}:  for each decimal value $dec[i]$ $(1 \leq i \leq n)$, two remainders $r[1,i]$ and $r[2,i]$, are computed such that 
\begin{center}
$r[1,i] = dec[i]\mod p_1 $ and $r[2,i] = dec[i]\mod p_2,\ 1 \leq i \leq n$
\end{center} 

\textbf{Step 5}: for each remainder $r[j,i],\ 1 \leq i \leq n,\ 1 \leq j \leq 2$, we compute positions $pos[1,1],\cdots, pos[2,n]$, as described by the following procedure:
\textit{
\begin{tabbing}
1.	l := 0; i := 1; n := dec.length; h := $2 \times p_2$;\\
2.	wh\= ile (i $\leq$ n) do\+ \\	
		be\= gin\+ \\
2.1.		fo\= r (j := 1 to 2) do\+ \\
				pos[$2 \times (i - 1) +j$] := l + (j - 1) + $2 \times r[j,i]$;\- \\
2.2.		l := l + h; \\
2.3.		i := i + 1;\- \\
		end;
\end{tabbing}
}

\textbf{Step 6}: sort the vector $pos$ in the ascending order;\\

\textbf{Step 7}: for each $pos[i], 1 \leq i \leq n\times 2 - 1$, insert one $A0$ at the $pos[i]^{th}$ between-character location of $f$. And, at the $pos[n\times 2]^{th}$ between-character location of $f$, insert two $A0$'s, to mark then end of the process.\\
\textbf{Time Complexity}: $O(n*k)$

\subsubsection{Recovery method}
\textbf{Input:} f: stego-PDF file, k: secret key \\
\textbf{Output:} s: secret message\\

\textbf{Step 1}:two co-primes $p_1, p_2$, are computed from $k$ such that,
\begin{center}
$p_1 = 2^{\lceil{\frac{k}{2}}\rceil}$; $p_2 = p_1 + 1$.
\end{center} 

\textbf{Step 2}: compute the block length in the file $f$ such that:
\begin{center}
$h = 2 \times p_2$
\end{center}

\textbf{Step 3}: retrieve the positions where $A0$'s have been inserted as described below:
\textit{
\begin{tabbing}
1.	j := 1; l := h; i := 1; count := 1; n := 0; exist := true;\\
2. get the first couple (a, b) of characters from $f$;\\
3.	wh\= ile (exist and !feof(f)) do\+ \\	
		be\= gin\+ \\
			if\=  (a != $A0$ and b = $A0$) then \+ \\
				be\= gin\+ \\
					pos[$i$] := count;\\
					i := i + 1;\- \\
				end; \- \\
			el\= se\+ \\
				if (a != $A0$ and b != $A0$) then do nothing; \\
				el\= se\+ \\ 
					if (a = $A0$ and b != $A0$) then count := count - 1;\\
					el\= se\+ \\  
						if (a = $A0$ and b = $A0$) then exist := false;\\
						end;\- \\
					endif;\- \\
				endif;\- \\
			endif;\\
			count := count + 1;\\
			c := the next character in f;\\
			a := b; \\
			b := c;\- \\
		end;\- \\
4.	n := (i - 1) / 2; 
\end{tabbing}
}

\textbf{Step 4}: Compute the remainders from the table $pos$ as follows:
\textit{
\begin{tabbing}
1.	l = 0; n = pos.length / 2;\\
2.	fo\= r (i := 1 to n) do\+ \\	
		be\= gin\+ \\
2.1.		fo\= r (j := 1 to 2) do\+ \\
				f := $[pos[(2 \times (i - 1)) + j] - (l + j - 1)] \mod 2$;\\
				r[f, i] := $[pos[(2 \times (i - 1)) + f] - (l + f - 1)] / 2$;\- \\
2.2.		l := l + h; \- \\
		end;
\end{tabbing}
}

\textbf{Step 5}: Compute each decimal value $dec[i]$ of a block of $s$ such that:
\begin{center}
$dec[i] = GarnerAlgorithm(\{p_1,p_2\}, \{r[1, i], r[2, i]\})$\\
$1 \leq i \leq n$.
\end{center}

\textbf{Step 6}: transform each $dec[i]$ in its binary sequence $sp[i,j],\ (1 \leq j \leq k)$ bits such that:
\begin{center}
$(dec[i])_2  =  \underbrace{sp[i, 1] sp[i, 2]\cdots sp[i, k]}_{k\ bits}$
\end{center}

\textbf{Step 7}: merge all the binary string into one, the secret $s$, such that:
\begin{center}
$ s[(i - 1)k + j] = sp[i, j],\ 1 \leq i \leq n, 1 \leq j \leq k$.
\end{center}
\textbf{Time Complexity}: $O(n*k)$

\subsubsection{Evaluation}
	In this approach, for each block of length $k$, 2 $A0$'s are inserted in the cover file. So to embed a full secret message $s$ divided into $n$ blocks of length $k$, $2n$ $A0$'s would be needed. Regardless the number of blocks we need to embed, an additional $A0$, would be added to allow the recovery method to stop when all the hidden bits have been recovered. The obtained result is resumed by the following theorem.
\begin{theorem}
.\\
Given a secret message $s$, a secret key $k$ such that number of blocks of length $k$, is given by $n = \frac{|s|}{k}$, and two primes $p_1, p_2$ such that $p_1 = 2^{\lceil{\frac{k}{2}}\rceil}$, $p_2 = p_1 + 1$, the number $N$ of $A0$'s insertions at between-character locations, to perform in a PDF file is given by:
\begin{center}
$N =
\begin{cases}
		3, \ if \ |s| \leq k \\
		2n+1, \ if \ |s| > k \\
\end{cases}$
\end{center}

\end{theorem}
$\blacksquare$
	
\subsection{Fourth Approach}
\	In this particular approach, there is no need of a secret key. Here, we embed only 3 $A0$'s, at 3 different positions in the cover file $f$. Their values, depend only on length of the secret message that a sender wants to send through a non-secure communication channel.
	
\subsubsection{Hiding method}
\textbf{Input:}  s: secret message; f: cover PDF file. \\
\textbf{Output:} f: cover PDF file with embedded \textit{A0}'s\\

\textbf{Step 1}: compute $n$, the length of the secret message $s$.

\textbf{Step 2}: insert one $A0$ at the $n^{th}$ between-character location in the file $f$.

\textbf{Step 3}: compute two co-primes $p_1, p_2$ such that,
\begin{center}
$p_1 = 2^{\lfloor \frac{n}{2} \rfloor}$; $p_2 = p_1 + 1$.
\end{center} 

\textbf{Step 4}: transform $s$ in its decimal value $dec$ such that,
\begin{center}
$dec = \sum_{i = 1}^n s[i] \times 2^{(n - i)}$.
\end{center}

\textbf{Step 5}:  compute two remainders $r[1]$ and $r[2]$ such that,
\begin{center}
$r[1] = dec\ mod\ p_1 $ and $ r[2] = dec\ mod\ p_2$.
\end{center} 

\textbf{Step 6}: for each remainder $r[i]\ (1 \leq i \leq 2)$, we compute positions $pos[1]$ and $pos[2]$ as follows:
\begin{center}
$pos[1] = n + 2*r[1]$, and
$pos[2] = n + 2*r[2] + 1$.
\end{center} 

\textbf{Step 7}: embed one $A0$ at $pos[1]^{th}$ and $pos[2]^{th}$ between-character locations in the file $f$.\\
\textbf{Time Complexity}:  $O(n)$

\subsubsection{Recovery method}
\textbf{Input:} f: stego-PDF file, \\
\textbf{Output:} s: secret message\\

\textbf{Step 1}: browse the stego-PDF file, until we cross the first $A0$, and store its position in $n$.

\textbf{Step 2}: compute two co-primes $p_1, p_2$ such that,
\begin{center}
$p_1 = 2^{\lfloor \frac{n}{2} \rfloor}$; $p_2 = p_1 + 1$.
\end{center} 

\textbf{Step 3}: browse the stego-PDF file, from the position $n$, until we cross the second $A0$, store its position in $pos[1]$ and the last $A0$, and store its position in $pos[2]$.

\textbf{Step 4}: permute if necessary the values of $pos[1]$ and $pos[2]$ as follows:
\textit{
\begin{tabbing}
be\= gin \+ \\
1. $pos[1]$ := $pos[1]$ - n;\\
2. $pos[2]$ := $pos[2]$ - n;\\
3. if $pos[1]$ is $odd$, permute with $pos[2]$;\- \\
end;
\end{tabbing}
}

\textbf{Step 5}: computes remainders $r[1]$ and $r[2]$ from positions $pos[1]$ and $pos[2]$ as follows:
\begin{center}
$r[1] = pos[1]/2$, and
$r[2] = (pos[2] - 1)/2$.
\end{center} 

\textbf{Step 6}: Compute the decimal value $dec$ such that:
\begin{center}
$dec = GarnerAlgorithm(\{p_1,p_2\}, \{r[1], r[2]\})$
\end{center}

\textbf{Step 7}: transform $dec$ in its binary sequence $s$ on $n$ bits length such that:
\begin{center}
$dec_2  =  \underbrace{s[1] s[2]\cdots s[n]}_{n\ bits}$
\end{center}
\textbf{Time Complexity}:  $O(n)$

\subsubsection{Evaluation}
	As with this method, we have the possibility to embed not more or less than 3 $A0$'s, no matter how long the message is, we've reached the following result.
\begin{theorem}
.\\
Given a secret message of length $n$ and two primes $p_1, p_2$ such that $p_1 = 2^{\lfloor \frac{n}{2} \rfloor}$ and $p_2 = p_1 + 1$. The number $N$ of $A0$'s insertions at between-character locations, to perform in a PDF file is given by:
\begin{center}
$ N = 3$
\end{center}
\end{theorem}

$\blacksquare$\\

The proof of this theorem is trivial, regarding the definition of the hiding method.

\section{Experimental results}

We conducted experiments on our approaches to make sure we reach our goal, which is to reduce the insertion of $A0's$ in a PDF file, to maintain a small difference between cover and stego PDF files, while increasing the amount of data that can be hidden in that PDF file serving as cover. \\

To have a better view of our results, we've chosen as inputs the following: secret message \textit{s = "This is a covert communication method."} (as in \cite{IW 10}), with $nchar = 38\ characters$ and a random PDF file. For that input \textit{I Shi et al.} inserted $247 A0's$ in a pdf file. As described by the following table. Note: $C$ is \textit{Character}, $F$ is \textit{Frequency}, $N$ is the number of $A0$'s for a character and $B$ is \textit{Bits}.\\

\begin{table}[h]
\begin{center}
 \begin{tabular}{*{2}{c}}
   \begin{tabular}{|*{4}{c|}}
     \hline
     \textbf{C} & \textbf{F} 	& \textbf{N} 		& \textbf{F*N} 					\\ \hline
     LF			& 1 			& 12 				& 12							\\ \hline
     \ 			& 5 			& 1    				& 5 							\\ \hline
     T 			& 1 			& 13 				& 13							\\ \hline
     a 			& 2 			& 7  				& 14							\\ \hline
     c 			& 3 			& 4  				& 12							\\ \hline
     d 			& 1 			& 14 				& 14							\\ \hline
     e 			& 2 			& 8  				& 16							\\ \hline
     h 			& 2 			& 9  				& 18							\\ \hline
     i 			& 4 			& 2  				& 8 							\\ \hline
   \end{tabular}
   &
   \begin{tabular}{|*{4}{c|}}
     \hline
     \textbf{C} & \textbf{F} 	& \textbf{N} 		& \textbf{F*N} 					\\ \hline
     m 			& 3 			& 5  				& 15							\\ \hline
     n 			& 2 			& 10 				& 20							\\ \hline
     o 			& 4 			& 3  				& 12							\\ \hline
     r 			& 1 			& 15 				& 15							\\ \hline
     s 			& 2 			& 11 				& 22							\\ \hline
     t 			& 3 			& 6  				& 18							\\ \hline
     u 			& 1 			& 16 				& 16							\\ \hline
     v 			& 1 			& 17 				& 17							\\ \hline
     \textbf{Total}	& 38		& \ 				& \textbf{247}					\\ \hline
   \end{tabular}
  \end{tabular}
 \end{center}
 \caption{Number of A0's inserted with the method of \textit{I Shi et al.}}
\end{table}

Regarding our methods, at the beginning we preprocessed the cover file, converted the secret message into its binary sequence, where each character was replaced by its ASCII code binary representation.\\
As we have 38 characters each represented on 8 bits, we would have 304 bits to hide in the cover PDF file. Let's assume $|s|$, the total number of bits and $bin$ the binary sequence of the secret message $s$.\\
\begin{table}[h]
\begin{center}
   \begin{tabular}{| l || c || r | }
     \hline
     \begin{tabular}{ l | c | r  }
     \textbf{C} 		& \textbf{H} 			& \textbf{ASCII Code} \\ \hline
     LF					& 0A 					& 00001010 \\ \hline
     \ 					& 20 					& 00100000 \\ \hline
     T 					& 54 					& 01010100 \\ \hline
     a 					& 61 					& 01100001 \\ \hline
     c 					& 63 					& 01100011 \\ \hline
     d 					& 64 					& 01100100 \\ \hline
     e 					& 65 					& 01100101 \\
   	 \end{tabular} 
     & 
     \begin{tabular}{ l | c | r  }
     \textbf{C} 		& \textbf{H} 			& \textbf{ASCII Code} \\ \hline
     h 					& 68 					& 01101000 \\ \hline
     i 					& 69 					& 01101001 \\ \hline
     m 					& 6D 					& 01101101 \\ \hline
     n 					& 6E 					& 01101110 \\ \hline
     o 					& 6F 					& 01101111 \\ \hline
     r 					& 72 					& 01110010 \\ \hline
     s 					& 73 					& 01110011 \\
   	 \end{tabular}
     & 
     \begin{tabular}{ l | c | r  }
     \textbf{C} 		& \textbf{H} 			& \textbf{ASCII Code} \\ \hline
     t 					& 74 					& 01110100 \\ \hline
     u 					& 75 					& 01110101 \\ \hline
     v 					& 76 					& 01110110 \\ \hline
     \  				& \  					& \ \\ \hline
     \  				& \  					& \ \\ \hline
     \  				& \  					& \ \\ \hline
     \  				& \  					& \ \\
   	 \end{tabular} \\
     \hline
   \end{tabular}
 \end{center}
 \caption{ASCII codes of the secret message's characters}
\end{table}\\
Where \textit{C} is \textit{Character}, \textit{H} is \textit{Hexadecimal} (the hexadecimal ASCII code of the character) and \textit{ASCII Code} is the binary ASCII code of the character

\subsection{First approach} 
 
To compute the the number $N$ of inserted $A0$'s we use Theorem 2, and thus we obtain to following results: $C$ is \textit{Character}, $F$ is \textit{Frequency} and $B$ is \textit{Bits}.\\

\begin{table}[h]
\begin{center}
 \begin{tabular}{*{2}{c}}
   \begin{tabular}{|*{4}{c|}}
     \hline
     \textbf{C} & \textbf{F} & \textbf{ASCII Code} 	& \textbf{B} 	\\ \hline
     LF			& 1 		& 00001010 				& 2							\\ \hline
     \ 			& 5 		& 00100000  			& 5 						\\ \hline
     T 			& 1 		& 01010100 				& 3							\\ \hline
     a 			& 2 		& 01100001  			& 6							\\ \hline
     c 			& 3 		& 01100011  			& 12						\\ \hline
     d 			& 1 		& 01100100 				& 3							\\ \hline
     e 			& 2 		& 01100101  			& 8							\\ \hline
     h 			& 2 		& 01101000  			& 6							\\ \hline
     i 			& 4 		& 01101001  			& 16 						\\ \hline
   \end{tabular}
   &
   \begin{tabular}{|*{4}{c|}}
     \hline
     \textbf{C} & \textbf{F} & \textbf{ASCII Code} 	& \textbf{B} 	\\ \hline
     m 			& 3 		& 01101101  			& 15						\\ \hline
     n 			& 2 		& 01101110 				& 10						\\ \hline
     o 			& 4 		& 01101111  			& 24						\\ \hline
     r 			& 1 		& 01110010 				& 4							\\ \hline
     s 			& 2 		& 01110011 				& 10						\\ \hline
     t 			& 3 		& 01110100  			& 12						\\ \hline
     u 			& 1 		& 01110101 				& 5							\\ \hline
     v 			& 1 		& 01110110 				& 5							\\ \hline
     \textbf{Total}	& 38	& \ 					& \textbf{136}				\\ \hline
   \end{tabular}
  \end{tabular}
 \end{center}
 \caption{Number of A0's inserted.}
\end{table}

In the column $B$, for each character we computed the number of bits having value 1 in its ASCII code, multiplied by the its frequency in the secret message $s$. Thus, one can see that:
\begin{itemize}
\item We've obtained a better result compare to results obtained with the method of \textit{I Shi et al.}: $N < 247\ A0$'s
\item We ensured the fact that the number of inserted $A0$'s is lower than the number of bits of $s$: $N < |s|$.
\end{itemize}

Note that the value $136$ represents the maximum number of $A0$'s that can be inserted in a cover PDF file, given the secret message taken as example in this study.

\subsection{Second approach} 

To compute the number $N$ of inserted $A0$'s, we use the Corollary 1, by replacing $|s|$ by its value and $k$ by its equation (\ref{E:mm9}). Thus:
\begin{center}
$N \sim
\begin{cases}
		\frac{3m}{ln(3m)} - \frac{2m}{ln(2m)} + 1, \ if \ |s| \leq \frac{17}{10}m \\
		(\frac{3m}{ln(3m)} - \frac{2m}{ln(2m)}) * n + 1, \ if \ |s| > \frac{17}{10}m \\
\end{cases}$
\end{center}

 And as the number of $A0$'s depends on $m$, we vary the value of $m$ to see where its optimal value stands. Here are some of the obtained results:\\
\begin{table}[h]
\begin{center}
 \begin{tabular}{*{3}{c}}
   \begin{tabular}{|*{5}{c|}}
   \hline
\textbf{m} 	& \textbf{k} 	& \textbf{t} 	& \textbf{n} 	& \textbf{n*t} \\ \hline
2	& 3		& 1		& 102	& 102 \\ \hline
12	& 20	& 3		& 16	& 48 \\ \hline
22	& 37	& 5		& 9		& 45 \\ \hline
32	& 54	& 6		& 6		& 48 \\ \hline
42	& 71	& 8		& 5		& 45 \\ \hline
52	& 88	& 10	& 3		& 30 \\ \hline
62	& 105	& 12	& 3		& 26 \\ \hline
72	& 122	& 13	& 3		& 39 \\ \hline
82	& 139	& 14	& 2		& 28 \\ \hline
   \end{tabular}
   &
    \begin{tabular}{|*{5}{c|}}
   \hline
\textbf{m} 	& \textbf{k} 	& \textbf{t} 	& \textbf{n} 	& \textbf{n*t} \\ \hline
92	& 156	& 16	& 2		& 32 \\ \hline
102	& 173	& 17	& 2		& 34 \\ \hline
112	& 190	& 18	& 2		& 36 \\ \hline
122	& 207	& 19	& 2		& 38 \\ \hline
132	& 224	& 21	& 2		& 42 \\ \hline
142	& 241	& 22	& 2		& 44 \\ \hline
152	& 258	& 23	& 2		& 48 \\ \hline
162	& 275	& 24	& 2		& 48 \\ \hline
172	& 292	& 25	& 2		& 50 \\ \hline
   \end{tabular}  
   &
   \begin{tabular}{|*{5}{c|}}
   \hline
\textbf{m} 	& \textbf{k} 	& \textbf{t} 	& \textbf{n} 	& \textbf{n*t} \\ \hline
179	& 304	& 25	& 1		& 25 \\ \hline
182	& 309	& 27	& 1		& 27 \\ \hline
192	& 326	& 28	& 1		& 28 \\ \hline
202	& 343	& 29	& 1		& 29 \\ \hline
212	& 360	& 30	& 1		& 30 \\ \hline
222	& 377	& 31	& 1		& 31 \\ \hline
232	& 394	& 32	& 1		& 32 \\ \hline
242	& 411	& 34	& 1		& 34 \\ \hline
252	& 428	& 35	& 1		& 35 \\ \hline				

   \end{tabular}
  \end{tabular}   
 \end{center}
 \caption{Number of $A0$'s ($N = n*t$), given the number of primes $t$ and number of blocks $n$, both obtained from $m$}
\end{table}

In this approach, the block length $k$ is not the secret key, but is computed from $m$ which is. And even the set of prime numbers used to compute remainders is generated from it.\\
By varying the different values of $m$, we came up we a certain number of curves.\\

\begin{figure}[h]
\centering
\includegraphics[scale=0.70]{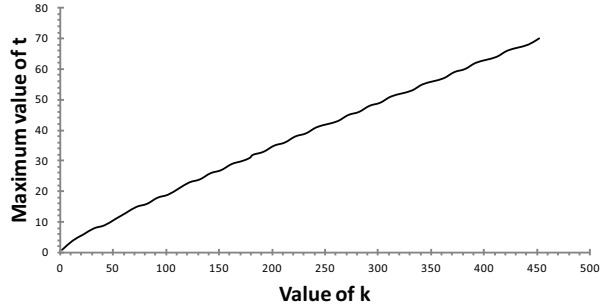} 
\caption{Evolution of the number $t$, of prime numbers with respect to $k$}
\end{figure}

This curve shows the growth of $t$ with respect to $k$ (or $m$). We can see that, the more $k$ grows, the more the number of prime numbers that would used in the computation of $A0$'s grows. And as each prime generates one $A0$, the number of $A0$'s grows too.\\

\begin{figure}[h]
\centering
\includegraphics[scale=0.70]{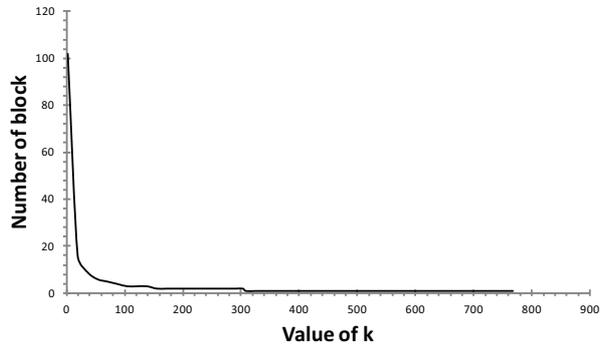} 
\caption{Evolution of the number of block with respect to $k$}
\end{figure}

Then, we generated a curve, showing that, the more $k$ gets closed to $|s|$, the more $n$, the number of blocks, decreases until it reaches the value 1; where it remains constant no matter the value $k$ (for $k < |s|$).\\

After having computed for each value of $m$, the block length $k$, the number of primes $t$ and the number of block $n$ of the secret message $s$, we generated a curve showing the growth of $N$ the number of $A0$'s that would be use to encode the secret message $s$, with respect to $k$ (or $m$).\\

\begin{figure}[h]
\centering
\includegraphics[scale=0.70]{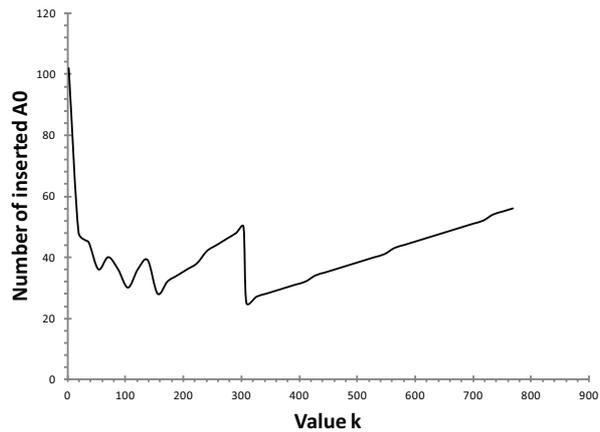} 
\caption{Evolution of the number of $A0$'s with respect to $k$}
\end{figure}

One can see that, when $k$ gets superior to $|s|$, $N$ the number of $A0$'s depends now on the number of primes $t$. Meaning that, the more $t$ grows the more the $N$ grows. Where $t$'s growth is a consequence of the growth of $s$, as shown by the first curve.\\

And for a value of $k$ taken between $1$ and $|s|$, the value of fluctuate, making it difficult to choose the right value of the key $m$, that lowers the number of inserted $A0$'s. But compare to the result of \textit{I Shi et al.} for a value of $k \in [1, |s|]$, the max value (this is when $k = 1$) is less than the half of value ($247 A0$'s) they've obtained.\\

Also, one can see that the optimal value of $N$ can be reached for $k \in [\frac{1}{4}|s|, \frac{3}{4}|s|]$. For that, $N < |s|$, and there are certain cases ($k \in [92, 102]$ and, $k \in [182, 222]$) where $N$ gets lower than the number of characters of the secret message $s$, which is hard to generalize.

\subsection{Third approach}

To compute the number $N$ of inserted $A0$'s, we use the Theorem 3, where :
\begin{center}
$N =
\begin{cases}
		3, \ if \ |s| \leq k \\
		2n+1, \ if \ |s| > k \\
\end{cases}$
\end{center}
And as the number of $A0$'s depends on $k$, we vary the value of $k$ to see where its optimal value stands. Here are some of the obtained results:
\begin{table}[h]
\begin{center} 
   \begin{tabular}{|*{2}{c|}}
     \hline
     \textbf{k} 	& \textbf{Value of N}	\\ \hline
     1 				& 609						\\ \hline
     2 				& 305 					\\ \hline
     3 				& 205 					\\ \hline
     16 			& 39 					\\ \hline
     $|s|$/4 		& 9 					\\ \hline
     $|s|$/2 		& 5 					\\ \hline
     3*$|s|$/4		& 5 					\\ \hline
     $|s|$ 			& 3 					\\ \hline
	 5*$|s|$/4		& 3 					\\ \hline
	 3*$|s|$/2 		& 3 					\\ \hline    
   \end{tabular}
 \end{center}
 \caption{Number of A0's with respect of k}
\end{table}

\begin{figure}[h]
\centering
\includegraphics[scale=0.70]{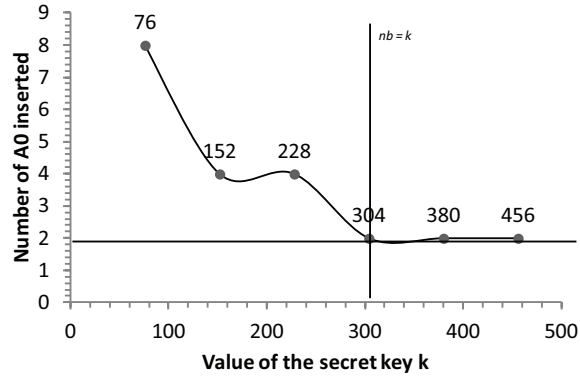} 
\caption{Evolution of the number of $A0$'s with respect to the key $k$}
\end{figure}
\ \\
From the above operations, whose some of the results are represented by the figure below, we can see that:
\begin{itemize}
\item For $k < 3$, $N > 247\ A0's > |s|$. Which is not a good situation;
\item For $k = 3$, $N = 205 < 247\ A0's$ and $N < |s|$. Meaning, from here we inserted less $A0$'s than with the method of \textit{I Shi et al.};
\item For $k \geq 16$, $N < 247\ A0's$ and $N < |s|$. From this point, $N$ starts to get lower than the number of characters of $s$. As for $k = 16$, we have $N = 39$, which is exactly the number of Characters contained in $s$.
\item For $152 \leq k < |s|$, $N = 5$. Meaning that at this point, the weight difference between the cover and the stego file is almost invisible;
\item For $k \geq |s|$, $N = 3$. $N$ remains constant no matter the value of the $k$.\\
\end{itemize}

So, to ensure that a minimum number of $A0$'s would be inserted in a cover PDF file, the sender and the receiver, should agree on a secret key with high value.\\

\subsection{Fourth approach}

First of all, compute the two remainders $p_1$, $p_2$ that would help us to compute positions.
\begin{itemize}
\item $p_1 = 2^{\lfloor \frac{304}{2} \rfloor} = 2^{152}$;
\item $p_2 = p_1 + 1 = 2^{152} + 1$.
\end{itemize}
Then, convert the $bin$ in its decimal value $dec$ and compute three positions where one $A0$ would be inserted in the PDF. Those positions are:
\begin{itemize}
\item First position: $pos[0] = |s| = 304$
\item Second position: $pos[1] = 2*(dec\ mod\ p_1) + |s|$
\item Third position: $pos[2] = 2*(dec\ mod\ p_2) + |s|$
\end{itemize}
Whatever the values of the computed positions, only 3 $A0's$ will be inserted. One can conclude that:
\begin{itemize}
\item The weight difference between the cover file and the stego file is 3 bytes;
\item The number $N$ of inserted $A0$'s is far smaller than the number inserted using \textit{I Shi et al.} method;
\item We ensured the fact that the $N \leq |s|$.
\end{itemize}

We can resume our results, for the chosen secret message of 38 characters, as shown by the following table :
\begin{table}[h]
\begin{center}
   \begin{tabular}{|*{6}{c|}}
   \hline
\textbf{ } 						& \textbf{I-Shi et al.} 	& \textbf{$1^{st}$ case} 	& \textbf{$2^{nd}$ case} 	& \textbf{$3^{rd}$ case} 	& \textbf{$4^{th}$ case}	\\ \hline
\textbf{N}			& 247			& 138			& $\in [25, 38[$				& $\geq$ 3 			& 3	\\ \hline
\textbf{files}		& 2				& 1				& 1								& 1 						& 1	\\ \hline
   \end{tabular}
 \end{center}
 \caption{Comparison of methods}
\end{table}

With these experiments we've shown the effectiveness and the correctness of our approaches.

\section{Discussion}

From the our results obtained, expressed in the previous section, we came up with some observations, regarding the choice of a secret key, to embed a secret message $s$, in a cover PDF file.\\

The number of signs that can be contained in a document page is closed to $1500$. Where a sign can be, space, punctuation, apostrophes, etc. Thus, the number of between-character locations in that page is close to $1500$ ($2^{10} < 1500 < 2^{11}$). It implies that:
\begin{itemize}
\item In the first and fourth approaches: for each $p_i$ multiple of $1500$, that is to say that $p_i = 1500*\alpha$ ($\alpha \geq 1$, $1 \leq i \leq 2$), we would need $\alpha$ page(s) to hide the number of $A0$'s generated by $p_i$.
\item In the second and third approaches: $h = t \times p_t$, where $h$ is the number of between character locations used to hide $A0$'s generated by $t$ prime numbers, and $t$ in the third approach equals 2. for $h$ multiple $1500$, that is to say that $h = 1500*\alpha$ ($\alpha \geq 1$), we would need $\alpha$ pages to hide a block of the secret message $s$.
\end{itemize} 

Thereby, the more $h$ or $p_i$ is high, the more we would need a cover PDF file with a high number of pages to embed our secret message. And here, the amount of embbedable information would depend on the approach selected for the purpose. Our approaches can be optimized even more, by using a compression algorithm on the secret message as done in \cite{IW 10}.\\

The advantage of our method is that it would be difficult to detect the integration of secret information in the cover file, while the inconvenient is that the file's number of pages can grow exponentially as it depends on $h$ or $p_i$.

\section{Conclusion}

A novel approach of PDF steganogaphy is proposed based on the Chinese Remainder Theorem. In this paper we presented four different techniques whose purpose is to increase the amount of information that can be hidden in a cover PDF file, while reducing considerably the number of $A0$'s insertions at between-character locations in that file, thus reducing the weight difference between a cover file and a stego file in which a secret message is embedded. We did this, by ensuring that the number of embedded $A0$'s would be less than the number of characters of $s$ or at least if $s$ grows higher, the number of inserted $A0$'s won't explode.
 Experimental results show the feasibility of the proposed methods and parameters to attain an optimal efficiency had been exposed. Further researches may be directed to improve these methods, and also to applying the data hiding scheme to other applications like watermarking for copyright protection, authentication of PDF files, etc.

\section{Acknowledgments}
This work was supported by {\it UMMISCO },  by {\it  LIRIMA } and by the {\it University of Yaounde 1}.

\end{document}